\documentclass[twocolumn,prl,showpacs,preprintnumbers,amsmath,amssymb,floatfix]{revtex4}
\pdfoutput=1
\usepackage{graphicx}
\usepackage{dcolumn}
\usepackage{bm}
\usepackage{verbatim}
\usepackage{tabularx}


\def\beq{\begin{equation}}
\def\eeq{\end{equation}}
\def\bea{\begin{eqnarray}}
\def\eea{\end{eqnarray}}

\newcommand{\gev}{\, {\rm GeV}}
\newcommand{\tev}{\, {\rm TeV}}

\def\textbfx#1{\section{#1}}

\def\g{\gamma}

\begin{document}

\preprint{CERN-PH-TH/2012-158}

\title{How well do we need to measure Higgs boson couplings?}

\author{Rick S. Gupta$^{a,b,c}$, Heidi Rzehak$^a\footnote{On leave from:
  Albert-Ludwigs-Universit\"at Freiburg, Physikalisches Institut,
  Freiburg, Germany.}$, James D. Wells$^{a,b}$}
\affiliation{
$^a$CERN Theoretical Physics, Geneva, Switzerland \\
$^b$Physics Department, University of Michigan, Ann Arbor, MI  USA \\
$^c$IFAE, Universitat Autonoma de Barcelona, 08193 Bellaterra, Barcelona, Spain
}

\date{\today}

\begin{abstract}
Most of the discussion regarding the Higgs boson couplings to Standard Model vector bosons and fermions is presented with respect to what present and future collider detectors will be able to measure. Here, we ask the more physics-based question of how well do we need to measure the Higgs boson couplings? We first present a reasonable definition of ``need" and then investigate the answer in the context of various highly motivated new physics scenarios: supersymmetry, mixed-in hidden sector Higgs bosons, and a composite Higgs boson. We find the largest coupling deviations away from the SM Higgs couplings that are possible if no other  state related to EWSB is directly accessible at the LHC. Depending on the physics scenario under consideration, we find targets that range from less than $1\%$ to $10\%$ for vector bosons, and from a few percent to tens of percent for couplings to fermions.

\end{abstract}

\maketitle


\textbfx{Introduction}
There are preliminary indications that a Higgs boson with mass of about $125\gev$ may have been seen in the data of the ATLAS and CMS experiments at LHC~\cite{ATLAS:2012ae,Chatrchyan:2012tx}.  If it is confirmed, the next question to ask is whether or not it is the SM Higgs boson. 
Establishing that the Higgs boson is a pure SM one is impossible, strictly speaking, but careful measurements of its couplings to the other Standard Model (SM) particles could confirm suspicions that it may not be a SM Higgs boson.  There are plenty of opportunities to check the couplings since a $125\gev$ SM Higgs boson has several substantive  
branching fractions~\cite{Dittmaier:2011ti}: $B(bb)\simeq 60\%$, $B(WW)\simeq 20\%$, $B(gg)\simeq 9\%$, $B(\tau\tau)\simeq 6\%$, $B(ZZ)\simeq 3\%$, $B(cc)\simeq 3\%$, etc. $B(\gamma\gamma)\simeq 0.2\%$ is also substantive due to the high mass resolution and relatively low background.

There are numerous studies that detail how well a collider can measure the Higgs boson couplings to the SM particles. These include LHC experiments in the near term~\cite{Rauch:2012wa}, capabilities of a high luminosity LHC upgrade~\cite{Klute:2012pu}, and the capabilities of high-energy $e^+e^-$ colliders, such as ILC~\cite{Abe:2010aa,Aihara:2009ad} and CLIC~\cite{Linssen:2012hp}.  It is appropriate that these studies have focused on achieving the best sensitivity possible on the Higgs boson couplings. However, what the studies typically do not address is, how well do we really need to know the couplings? Our aim here is to answer this question in the context of three of the most highly motivated ideas of physics beyond the SM. The results
will not be generic for any conceivable physics-beyond-the-SM
scenario, but some genericness issues among these ideas will be
addressed in the conclusions.

To answer this question we must first decide on the criteria for ``need". There are many possibilities. One answer is that we must do what it takes to get measurements down to at least as good as the theory errors of impossible or difficult to compute higher-order corrections. For example, relating the bottom quark partial width to the bottom quark Yukawa coupling requires a theory calculation to compare with the experimental measurement. It is impossible to pin down the bottom quark Yukawa coupling to better than a few percent no matter how well the partial width is measured due to the higher order corrections. Thus, measuring the partial widths, or equivalently measuring some ``observable coupling", to significantly better than a few percent is not needed, although such precision would be welcome for future generations who might be able to calculate better than us.

Another way to answer the question is, how well do we have to measure the Higgs couplings to see that it is exotic (i.e., non-SM) despite seeing no other ``Higgs sector" or ``symmetry breaking sector" state directly at the LHC. If we do see such states, we already know the Higgs sector is exotic, and so a coupling deviation is not qualitatively surprising or illuminating, although of course it would be quantitatively interesting. Operationally, this definition of need means that we must find the largest coupling deviations away from the SM Higgs couplings that are possible if no other ``Higgs sector" or ``symmetry breaking sector" state is directly accessible at the LHC.  This nonobservation criteria will be specified for each of the
beyond-the-SM physics scenarios, relying on previous studies. In
general, we consider at least 100 ${\rm fb}^{-1}$ of integrated
luminosity at full LHC design center-of-mass energy of 14 TeV.

There is no fully model-independent analysis to this determination, and so we shall give the answer within three different contexts. The first context is a singlet Higgs boson mixed in with the SM Higgs boson. The second context is a composite Higgs boson. And the third context we shall study is the Higgs bosons within the Minimal Supersymmetric Standard Model (MSSM). Our results will be summarized and conclusions given in the final concluding paragraphs.

\textbfx{Mixed-In Singlet Higgs Boson}
Let us first consider a theory where there is an exotic Higgs boson that is a singlet under the SM gauge group but may spontaneously break symmetries in some hidden sector group. Since this extra field $\Phi_S$ is a scalar and gets a vacuum expectation value it can mix at the renormalizable level with the SM Higgs doublet $H_{SM}$ through the  operator $|H_{SM}|^2|\Phi_S|^2$~\cite{Wells, Bowen:2007ia}. There will be two resulting CP-even mass eigenstates, which are neither purely SM nor purely hidden sector. 
The mixing angle $\theta_h$ that takes the fields from gauge eigenstates to mass eigenstates is a new crucial parameter for Higgs boson phenomenology. 
The couplings of the two Higgs bosons $h$ and $H$ to fermions and gauge bosons, with respect to the SM Higgs boson are,
\bea
g_h^2 =c_h^2~g^2_{{SM}} \\
g^2_H= s_h^2~g^2_{{SM}}.
\eea
where $c_h=\cos \theta_h$ and $s_h=\sin \theta_h$. 
We will choose $h$ to be the SM-like Higgs boson and hence $s_h^2<0.5$. 

We want to consider a scenario where $\theta_h$ and $m_H$ are such that the non SM-like Higgs boson  is not discovered even after data is collected by the 14 TeV LHC. For a given value of $\theta_h$ the deviation in the couplings would be,
\beq
\Delta g_{h}/g_{{SM}} \approx -s^2_h/2
\label{acc}
\eeq
where $\Delta g_{h} = g_h - g_{{SM}}$.
This is the required measurement accuracy for a given value of $\theta_h$. We want to find the maximum value of $s_h^2$ for which the heavier Higgs boson, $H$, with a given $m_H$ is  not detectable, and all other constraints, such as precision electroweak constraints, are satisfied. Through the relation in eq.~\ref{acc}, this will give us the minimum accuracy of measurement required to have a chance to detect deviations in Higgs couplings in such a scenario.  

\begin{figure}[t]
\includegraphics[width=0.9\columnwidth]{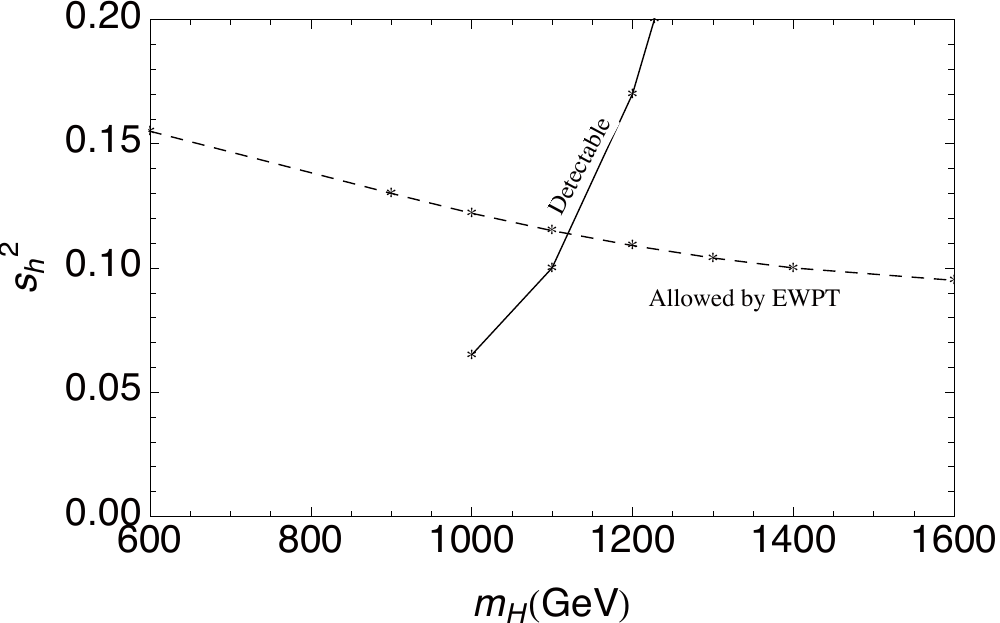}
\caption{
We show above the area in the $s_h^2-m_H$ plane  allowed by electroweak precision tests at the 90$\%$ CL in the presence of a mixed-in singlet Higgs boson. We also show the detectability curve (solid line) above which the scalar $H$ is detectable with 100 fb$^{-1}$ data at the 14 TeV LHC.  The maximum allowed $s_h^2$-value that can both evade detection and be consistent with precision electroweak constraints is thus given by the intersection of the two lines and is $s_h^2=0.12$.}
\label{fig:mixedinsinglet}
\end{figure}

In Ref.~\cite{Bowen:2007ia}, it was shown that for $m_H=1.1$ TeV and $s_h^2=0.1$ the scalar $H$ is barely detectable with only 13 signal events, in $100\, {\rm fb}^{-1}$ data, versus 7 background events.  For higher values of $m_H$,  higher values of $s_h^2$ are required to give the same number of events. We show this by the `Detectability Curve' in Fig.~\ref{fig:mixedinsinglet} where we have plotted values of $s_h^2$ that give the same cross-section as the production of a 1.1 TeV $H$ with $s_h^2=0.1$. As $s_h^2$  rises, the width of $H$ would also increase so that the number of signal events in any given mass window would decrease. This effect is not incorporated in the detectability curve in  Fig.~\ref{fig:mixedinsinglet} and including this effect will give an even steeper slope for  $m_H>1.1$ TeV. For the Higgs production cross-section we have used the lowest order expressions given in  
Ref.~\cite{hep-ph/9504378} and the MSTW parton density functions~Ê\cite{mstw}.

 Too high  values of $s_h^2$ are already ruled out by electroweak precision data. This is also shown in Fig.~\ref{fig:mixedinsinglet}. Up to one loop level, the contribution to the electroweak parameters $S$ and $T$ are given by the expressions,
 \bea
S = c_h^2 S_{SM}(m_h)+ s_h^2 S_{SM}(m_H) \\
T = c_h^2 T_{SM}(m_h)+ s_h^2 T_{SM}(m_H) 
 \eea
 where $S_{SM}$ and $T_{SM}$ are the contributions of the SM Higgs boson. We have used the one loop expressions that appear in Ref.~\cite{hep-ph/9409380}. To compute this bound we have taken $m_h=125$ GeV and  imposed the requirement that the contributions  to the $S$ and $T$ parameters are within the 90$\%$ CL $S-T$ contour in Ref.~\cite{723875} where $U$ is appropriately fixed to $0$. Another upper bound on $s_h^2$ comes from perturbative unitarity constraints for the scalar $H$. We have checked that these constraints are much weaker than precision constraints.

 It is clear from Fig.~\ref{fig:mixedinsinglet} that the maximum value of $s_h^2$ allowed by precision tests and such that the 14 TeV LHC barely misses the non-SM like Higgs boson $H$ even with 100 fb$^{-1}$ data is $s_h^2=0.12$. Using eq.~\ref{acc} we see that this corresponds to,
 \beq
 (\Delta g_{h}/g_{{SM}})^{target}\approx -6 \%.
 \eeq
 This is the physics target for measurement of the Higgs couplings in this scenario with the mixed in singlets.  This target value is equally applicable for Higgs couplings to any of the SM particles: $h\bar bb$, $h\bar tt$, $h\tau^+\tau^-$, $hW^+W^-$, $hZZ$, $h gg$, $h\g\g$ etc. 
 
At first look, the LHC is unlikely to ever get to the $6\%$ sensitivity. However, given that a $125\gev$ Higgs boson has many potentially detectable final states, 
it would be interesting to study what is the sensitivity  achievable under the assumption that all couplings are uniformly suppressed, as is the case with this mixed-in singlet example. The answer would surely be better than the sensitivities quoted for each individual final state; however, it is unlikely that it could reach as low as $6\%$ required here.

\textbfx{Composite Higgs boson}
We want to consider composite Higgs models where the Higgs boson is a pseudo-Goldstone boson and thus its mass is much lighter than the strong scale. Explicit models realizing this are Little Higgs models~\cite{ArkaniHamed:2002qy} and Holographic composite Higgs models~\cite{Contino:2003ve,Agashe:2004rs}. 
An effective field theory for such a strongly interacting light Higgs (SILH) boson has been developed in Ref.~\cite{Giudice:2007fh}. The SILH lagrangian contains higher dimensional operators involving SM fields that supplement the SM lagrangian. It is characterized by two independent parameters: the mass of the new resonances $m_\rho$ and their coupling $g_\rho$. 
The decay constant $f$, which is analogous to the pion decay constant $f_{\pi}$, is given by,
\beq
m_\rho=g_\rho f
\eeq
where $g_\rho \leq4 \pi$.

Here we do not list all the operators in the SILH lagrangian but only those relevant to us, i.e those that affect the Higgs couplings in the leading order or those that constrain $m_\rho$, 
\bea
{\cal L}_{SILH}& =&  \frac{c_{H}}{2f^2} \partial^\mu (H_{SM}^\dagger H_{SM}) \partial_\mu  (H_{SM}^\dagger H_{SM}) \nonumber \\&&+\frac{c_y y_f}{ f^2} H_{SM}^\dagger H_{SM} \bar{f}_L {H_{SM}} f_R
\nonumber\\&&+ \frac{c_S g g'}{4 m^2_\rho}(H_{SM}^\dagger \sigma_I H_{SM})B_{\mu \nu}W^{I\mu \nu} +h.c. ..
 \eea
where $y_f$ is the Yukawa coupling of fermion $f$ to the Higgs boson, $g$ and $g'$ are the $SU(2)$ and the $U(1)$ gauge couplings, and $\sigma_I$ the Pauli matrices. 
$H_{SM}$, $f_L$, $f_R$, $B_{\mu \nu}$ and $W^{I \mu \nu}$ denote the Higgs doublet, the left-handed and right-handed fermion
fields and the $U(1)$ and $SU(2)$ gauge field strength, respectively.  
The coefficients of the above operators have been estimated using Naive Dimensional Analysis (NDA)~\cite{Manohar:1983md, Giudice:2007fh} such that the couplings $c_H$, $c_y$ and $c_S$ are expected to be ${\cal O}(1)$ numbers.  Note that the operator with the coupling $c_S$ does not appear in the list in Ref.~\cite{Giudice:2007fh} as a different basis has been used in Ref.~\cite{Giudice:2007fh}.  The coupling $c_S$ is a linear combination of the couplings $c_W$ and $c_B$ in Ref.~\cite{Giudice:2007fh}. The operators with coefficients $c_H$ and $c_y$ lead to the leading deviations in Higgs couplings with respect to the SM,
\bea
\frac{\Delta g_V}{g_{V}^{SM}}&=& -c_H \xi/2 + \dots\label{vec}\\
\frac{\Delta g_f}{g_f^{SM}}&=& -c_H \xi/2-  c_y \xi + \dots \label{ferm}\\
\frac{\Delta g_{g}}{g_{g}^{SM}}&=& -c_H \xi/2-c_y \xi + \dots \label{glu}\\
\frac{\Delta g_{\g}}{g_{\g}^{ SM}}&=& -c_H \xi/2- \frac{c_y \xi}{1+J_\g(m_H^2)/I_\g(m_H^2)} + \dots \nonumber\\
&=&-c_H \xi/2+0.3~{c_y \xi} + \dots\label{cd}
\eea
where $\xi =v^2/f^2=g_\rho^2 v^2/m_\rho^2$ and $g_V, g_f, g_g$ and $g_\g$ are the Higgs couplings to gauge bosons, fermions, gluons and photons, respectively. 
$\Delta g_X$ denotes the difference between the coupling $g_X$ and $g_X^{SM}$ with $X = V,\, f,\, g,\, \g$ where ${g_{g}^{SM}}$ and $g_{\g}^{ SM}$ are 
loop-induced couplings.
The vacuum expectation value $v$ is $v \simeq 246$~GeV.
We have kept terms only up to first order in $\xi$.
 In the last equation, $I_\g$ and $J_\g$ are functions related to the top and $W$-loops in  $h\g \g$ diagrams whose explicit forms can be found in Ref.~\cite{Giudice:2007fh}.  In the second line of the same equation we have substituted the values of $I_\g$ and $J_\g$ taking $m_h=125$ GeV.  
For phenomenologically relevant cases it has been shown in Ref.~\cite{Low:2009di} that $c_H$ is always positive (an exception are models in the presence 
of a doubly charged scalar field) so that this operator always leads to suppression of composite Higgs couplings with respect to the SM. 
Note that for the $hgg$ and $h \g \g$  couplings (i.e $g_g$ and $g_\g$), the respective contributions from the operators, $({H_{SM}}^\dagger H_{SM}) G^{I\mu \nu} G^I_{\mu \nu}$ and $(H_{SM}^\dagger H_{SM}) F^{\mu \nu} F_{\mu \nu}$, $G^{I\mu \nu}$ and $F_{\mu \nu}$ being the gluon and the photon field strength, are sub-dominant, 
 as they are suppressed respectively by   $y_t^2/g^2_\rho$ and $g^2/g^2_\rho$ factors~\cite{Giudice:2007fh}. 

Now let us look at existing constraints and future LHC reach for the above parameters. The coupling  $c_S/m^2_\rho$ above is proportional to the precision electroweak parameter $S$.  From the constraints on the $S$-parameter, we can derive the following constraint on $m_\rho$~\cite{Giudice:2007fh},
 \beq
m_\rho \gtrsim 3{~\rm TeV.}
\label{mrho}
\eeq
Note that the constraint from the $T$-parameter is more severe but this is avoided by imposing custodial symmetry in specific  composite Higgs models. There is another contribution to precision observables due to the fact that the cancellation of divergences between the Higgs and gauge boson contributions that takes place in the SM,
 no longer occurs for a composite Higgs boson with reduced couplings to the gauge bosons. This leads to logarithmically divergent contribution to precision observables~\cite{Espinosa:2010vn}. 
 The constraint due to this effect has been evaluated in Fig.1.14  in Ref.~\cite{cdr} at the  99$\%$ CL. At   90$\%$ CL the same calculation gives  the constraint~\footnote{We thank A. Thamm for confirming this.},
\beq
c_H \xi = c_H \frac{g^2_\rho v^2}{m^2_\rho}\lesssim 0.15.
\label{es}
\eeq
taking $m_H= 125$ GeV.
Direct LHC probes are expected to be much less sensitive than existing precision constraints.  
With 300 fb$^{-1}$ data   diboson  ($WW$) production  in vector boson fusion (VBF) processes is expected to be sensitive only for $c_H \xi >0.5$ and double Higgs production is expected to be sensitive  for $c_H \xi \approx 1$~\cite{Contino:2010mh}. Production of strong resonances  is not expected to be competitive with precision constraints  for the higher $\xi$ values relevant to us (see Fig.1.14  in~\cite{cdr} and Ref.~\cite{appear}).
 
\begin{figure}[t]
\includegraphics[width=0.9\columnwidth]{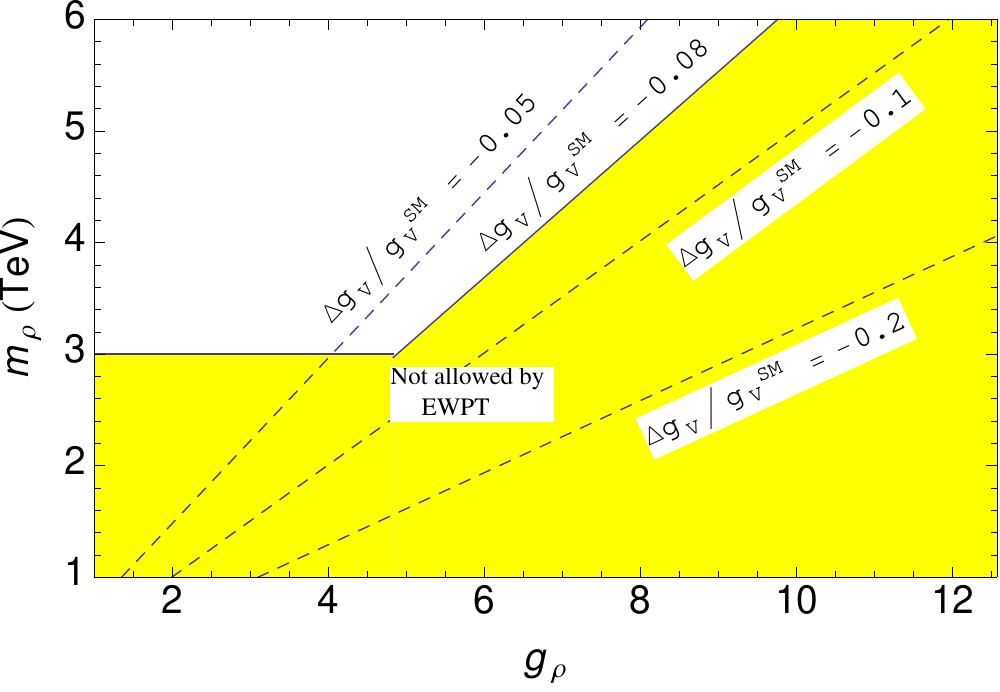}
\caption{Deviations  of  composite Higgs couplings to vector bosons from SM values ($\Delta g_V/g_V^{SM}$) for $c_H=1$. Note that whereas we have given the coupling deviations only to first order in $\xi$ in eq.~\ref{vec}, in this figure we have used the expression  $(1/\sqrt{1+c_H \xi}-1)$ for the coupling deviation, still neglecting terms suppressed by
$g/g_\rho$.
\label{fig:SILH}}
\end{figure}

We  first plot the fractional deviation in the gauge boson couplings, $\Delta g_V/g_V^{SM}$,  in Fig.~\ref{fig:SILH}. We have also marked the areas ruled out by  constraints. Note that  we have made the above plot for $c_H=1$ but  the values for the target derived would  not depend on $c_H$. This is because the  condition that goes into determining the coupling target  is eq.~\ref{es} which puts an upper bound, ${(c_H \xi)}_{max}=0.15$. This upper bound puts,  irrespective of the value of $c_H$, an upper bound on the coupling deviations in eq.~\ref{vec}, which are functions of the product $c_H \xi$.  We find the target to be,
\beq
(\Delta g_V/g_V^{SM})^{target} \approx-(c_H\xi)_{max}/2\approx-0.08.
\eeq
The target for the $hff$ and $hg g$ couplings  in eq.~\ref{ferm} and eq.~\ref{glu}, depends also on $c_y$, and is,
\beq
- (c_H\xi)_{max}/2- \frac{c_y}{c_H} (c_H \xi)_{max}\approx-0.08-0.15 \frac{c_y}{c_H}. 
\eeq
Although we have not focussed on $h\gamma\gamma$ coupling determinations, because in other scenarios they are derived directly from the $hWW$ and $h\bar tt$ couplings, we nevertheless give the simplified expression for its deviation within this composite Higgs scenario. Using eq.~\ref{cd} we find that $(\Delta g_\g/g_\g^{SM})^{target}$ is
\beq
- (c_H\xi)_{max}/2 +0.3 \frac{c_y}{c_H} (c_H \xi)_{max}\approx-0.08+0.05 \frac{c_y}{c_H}. 
\eeq

Thus if the parameter $c_y/c_H$ is known in a theory, the physics target for the above couplings can be found. 
 We can summarize the results by stating that the target for Higgs couplings to vector bosons in the composite Higgs model is about $8\%$, while for coupling to fermions it is tens of \%, depending on the unknown value of the ratio $c_y/c_H$.

\textbfx{MSSM Higgs bosons}
The MSSM~\cite{Martin:1997ns} has two Higgs doublets that mix. 
Typically there is one CP-even mass eigenstate $h$ that stays light and SM-like
and there is a full doublet of states ($H, A, H^\pm$) that is nearly degenerate and much heavier.  
The question we ask is what is the largest deviation of the couplings of $h$ compared to the SM Higgs boson in the circumstances where the other heavier Higgs bosons remain undiscovered at the LHC? In the case of the MSSM, there is the possibility of  discovering superpartners even if only a single CP-even Higgs boson is discovered. Although the superpartners are not strictly states related directly to EWSB, their presence would imply the presence of other Higgs bosons. Hence another important question to ask would be: what is the largest deviation of the couplings of $h$ compared to the SM Higgs boson in the circumstances where the other heavier Higgs bosons and the superpartners of SM particles, remain undiscovered at the LHC? We will answer both these questions in this section.

The MSSM mass matrix for the CP-even scalars can be expressed in terms of two parameters at tree level, 
the mass of the CP-odd Higgs boson $m_A$ and $\tan \beta$, which is the ratio of vevs $v_u/v_d$ of the two Higgs doublets. A $2\times 2$ matrix diagonalizes the Higgs bosons from the $\{ H^0_d,H^0_u\}$ basis to the mass eigenvalue basis $\{ h,H\}$ by the mixing angle $\alpha$. At tree-level the lightest eigenvalue is $m_Z^2\cos^22\beta$, when $m_A\gg m_Z$
with $m_Z$ being the Z~boson mass,  but can be higher through radiative correction contributions $\Delta_{ij}$ that are added to the mass matrix. The couplings of the light SM-like Higgs boson are related to the SM Higgs couplings by trigonometric factors of $\alpha$ and $\beta$:
\bea
\frac{g_u}{g^{SM}_u}=\frac{\cos\alpha}{\sin\beta}, ~~
\frac{g_d}{g^{SM}_d}= \frac{-\sin\alpha}{\cos\beta}, ~~
\frac{g_V}{g^{SM}_V}= \sin{(\beta-\alpha)}. \nonumber
\eea

\begin{figure}[t]
\includegraphics[width=0.9\columnwidth]{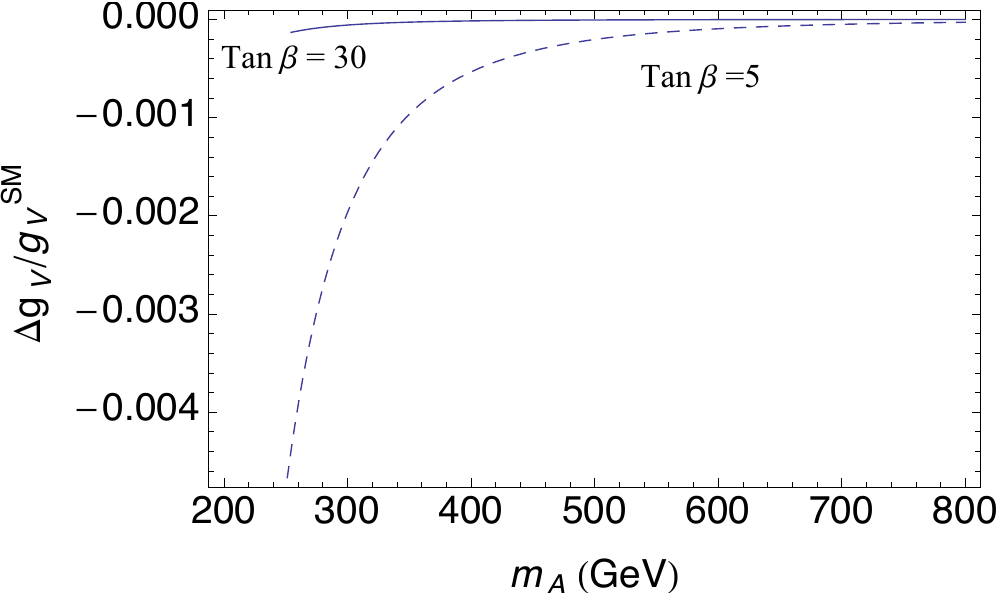}
\caption{ $\Delta g_V/g_V^{SM}$ as a function of $m_A$ with $\Delta_{22}$, the  radiative correction to the ${\cal M}^2_{22}$ entry of the Higgs mass matrix,   chosen to obtain $m_h=125\gev$. Other values of $\Delta_{ij}=0$. For the solid line we have taken $\tan \beta =30$ and for the dashed line $\tan \beta =5$.}
\label{fig:rv}
\end{figure}

\begin{figure}[t]
\includegraphics[width=0.9\columnwidth]{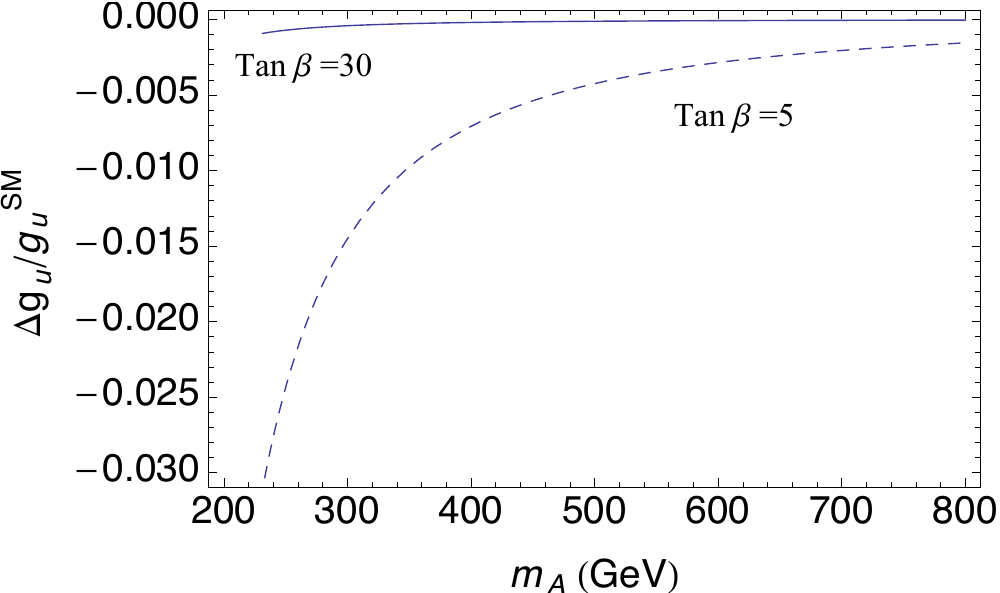}
\caption{We plot  $\Delta g_u/g_u^{SM}$ as a function of $m_A$ with $\Delta_{22}$, the  radiative correction to the ${\cal M}^2_{22}$ entry of the Higgs mass matrix,   chosen to obtain $m_h=125\gev$. Other values of $\Delta_{ij}=0$.  For the solid line we have taken $\tan \beta =30$ and for the dashed line $\tan \beta =5$.}
\label{fig:ru}
\end{figure}

\begin{figure}[t]
\includegraphics[width=0.9\columnwidth]{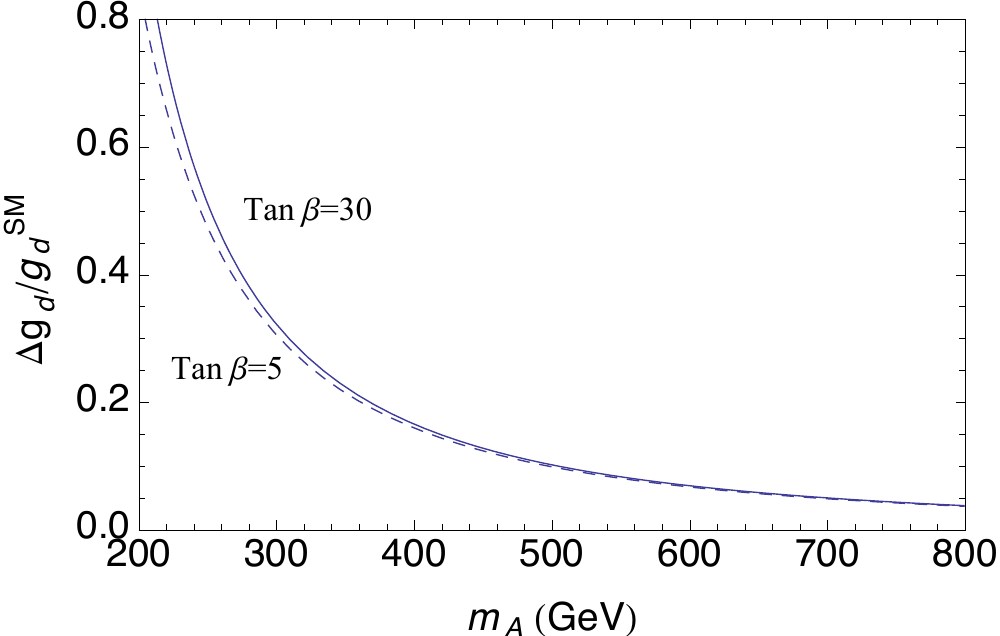}
\caption{We plot  $\Delta g_d/g_d^{SM}$ as a function of $m_A$ with $\Delta_{22}$, the  radiative correction to the ${\cal M}^2_{22}$ entry of the Higgs mass matrix,   chosen to obtain $m_h=125\gev$.  Other values of $\Delta_{ij}=0$.  For the solid line we have taken $\tan \beta =30$ and for the dashed line $\tan \beta =5$.}
\label{fig:rd}
\end{figure}

To achieve $125\gev$ there must be substantive radiative corrections to the light-Higgs mass. The dominant correction is generally to the ${\cal M}^2_{22}$ element of the mass matrix 
due to the large top Yukawa coupling and $H^0_u$ coupling to the top quarks.
 Thus, to start with,  the only modification we make of the tree-level matrix is to add a sufficiently large correction $\Delta_{22}$ to the mass matrix to obtain $m_h=125\gev$, while keeping the other $\Delta_{ij}=0$. From $\tan\beta$ and the mixing angle $\alpha$ we can find the deviations in the couplings, $
\Delta g_i/g_i^{SM} $. In Fig.~\ref{fig:rv},~\ref{fig:ru} and~\ref{fig:rd} we plot $\Delta g_V/g_V^{SM} $,  $\Delta g_u/g_u^{SM} $  and $\Delta g_d/g_d^{SM} $as a function of $m_A$ for two different values of $\tan \beta$. We see that the maximum deviations occur in the couplings with the down-type quarks. The deviations of the Higgs boson coupling to SM up-quarks and especially vector boson is typically small, which is well known in the literature~\cite{HaberOthers}. Therefore,  the target values for up-quark ($\sim {\rm few}\%$) and vector boson ($< 1 \%$) couplings are much smaller than those of down-type quarks.


For large $\tan\beta$ and also for non-typical hierarchies chosen for supersymmetry partner masses, one can find substantive contributions to other $\Delta_{ij}$ radiative correction entries to the Higgs boson mass matrix in addition to just $\Delta_{22}$.  Furthermore, one can also find almost any hierarchy among the $\Delta_{ij}$ values, although having $\Delta_{1j}\gg \Delta_{22}$ for $j=1$ or $2$ is generally not expected. We therefore investigate the corrections to the Higgs boson couplings to SM states under various choices of $\Delta_{ij}$ subject to the constraint that one SM-like eigenvalue must be at $125\gev$. There is also the potential for sizable finite $b$-quark mass corrections. However, those arise from lighter superpartners that come in loops, and for now we neglect those contributions with the assumption that they are decoupled effects compared to the Higgs mass matrix terms. Later, when we discuss a specific supersymmetry scenario, and precise superpartner masses are computed, we will include this effect.

In the MSSM the leading contributions~\cite{Djouadi:2005gj} to the radiative corrections to the different matrix elements are
\bea
\Delta_{22}&\propto& 24 \,y^4_tv^2 \log M_s/m_t +y_t^4v^2x_ta_t(12-x_ta_t)+\cdots \nonumber \\
\Delta_{11}&\propto& - y^4_tv^2 ~ \bar{\mu}^2x_t^2 +\cdots \nonumber \\
 \Delta_{12}&\propto& -y^4_tv^2 ~ \bar{\mu} x_t (6-a_t x_t) +\cdots \nonumber 
 \eea
where $\bar\mu=\mu/M_s$, $a_t=A_t/M_s$ and $ x_t=X_t/M_s=(A_t- \mu \cot \beta)/M_s$ with $X_t$ being the stop mixing parameter. $\mu$ denotes the Higgs 
superfield mixing parameter, $A_t$ the trilinear coupling of the top sector and $y_t$ the top Yukawa coupling. $M_s$ is defined as the arithmetic average of the stop masses, $M_s = 1/2\, (m_{\tilde{t}_1} + m_{\tilde{t}_2})$. For no mixing, $x_t=a_t-\mu \cot \beta=0$  which implies that  $\Delta_{11} \approx 0$ and $\Delta_{12} \approx 0$. For maximal mixing $x_t = \sqrt{6}$. Also note that for large $\tan \beta$, $x_t \approx a_t$ so that $6-x_t a_t \approx 0$ for maximal mixing,  which gives  $\Delta_{12} \approx 0$ even in this case for large $\tan \beta$. 
\begin{figure}[t]
\includegraphics[width=0.9\columnwidth]{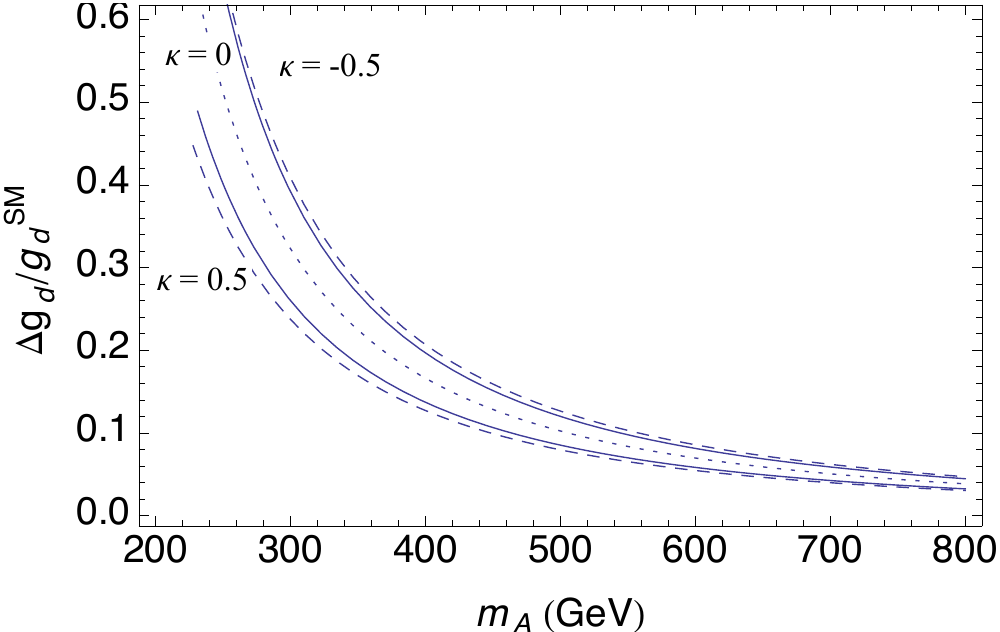}
\caption{ $\Delta g_d/g_d^{SM}$ as a function of $m_A$ for various values of $\kappa$ where $\Delta_{11}=\kappa \Delta_{22}$, and  $\Delta_{12}=0$. The overall contribution due to  radiative corrections has been chosen such that we get $m_h=125\gev$.  For the solid line we have taken $\tan \beta =30$ and for the dashed line $\tan \beta =5$.}
\label{fig:kappa}
\end{figure}

\begin{figure}[t]
\includegraphics[width=0.9\columnwidth]{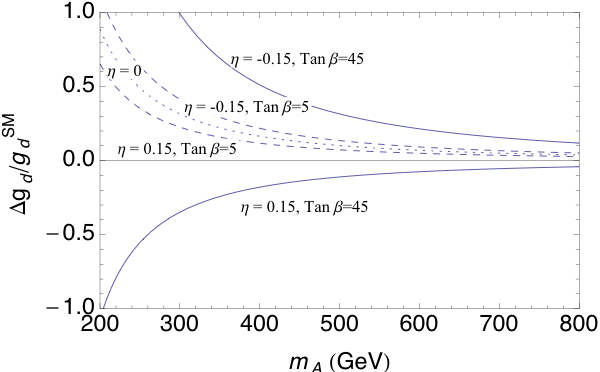}
\caption{ $\Delta g_d/g_d^{SM}$ as a function of $m_A$ for $\Delta_{11}=0$ and various values of $\eta$, where $\Delta_{12}=\eta \Delta_{22}$. The overall contribution due to  radiative corrections has been chosen such that we get $m_h=125\gev$.  }
\label{fig:eta}
\end{figure}

To see how non-zero values of $\Delta_{11}$ and $\Delta_{12}$ affect our results we first take $\Delta_{11}=\kappa \Delta_{22},  \Delta_{12}=0$ and plot the deviations in down-type quark couplings.  
This is shown in Fig.~\ref{fig:kappa} which shows a rapid convergence of the $h$ coupling to the SM value for large $m_A$ irrespective of the value of $\tan \beta$. 
Next we take $\Delta_{11}=0, \Delta_{12}=\eta \Delta_{22}$ and plot again the down-type quark couplings in Fig.~\ref{fig:eta}.  We find that in this case bigger deviations are possible especially for large $\tan \beta$~\cite{Loinaz:1998ph}, even with rather small values of $\eta$. This can be demonstrated analytically by computing the deviations in the down-type quark couplings for $m_A \gg m_Z$~\cite{Djouadi:2005gj},
\beq
\frac{g_d}{g^{SM}_d}\approx 1+ \frac{2 m_Z^2}{m_A^2}- \frac{\Delta_{12}}{m_A^2} \tan \beta.
\label{offd}
\eeq
Note, in particular, that the deviations grow with $\tan \beta$.

\begin{figure}[t]
\includegraphics[width=\columnwidth]{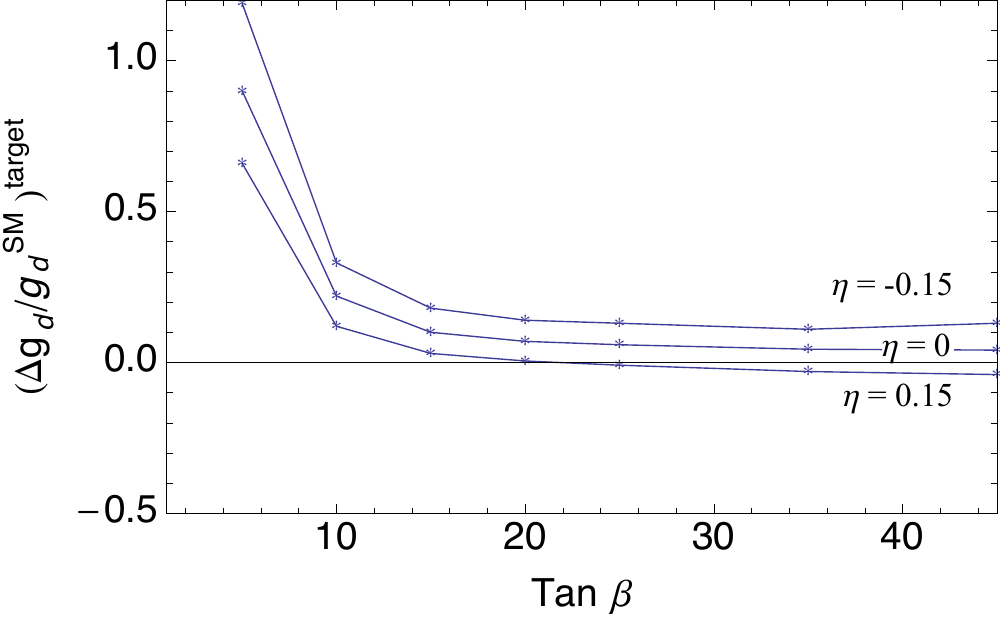}
\caption{We plot the physics target  $(\Delta g_d/g_d^{SM}) ^{target}$ as a function of $\tan \beta$ for $\Delta_{11}=0$ and different $\eta$ values, where $\Delta_{12}=\eta \Delta_{22}$ . The overall contribution due to  radiative corrections has been chosen such that we get $m_h=125\gev$. 
The target is the maximum deviation in $\Delta g_d/g^{SM}_d$ when no other Higgs state  is detectable.}
\label{fig:rdt}
\end{figure}

To find the physics target for Higgs coupling determination we need to know how well the LHC will be able to detect heavy Higgs bosons of supersymmetry. We model this after Fig.~1.21 of Ref~\cite{cdr},
which was gleaned from  Chapter 19 in Ref.~\cite{:1999fr},
which plots the minimum value of $m_A$ such that only a single light CP-even Higgs boson and no other Higgs boson is detectable at the LHC with 300 fb$^{-1}$ data for a given $\tan \beta$. The $\Delta g_d/g_d^{SM} $ corresponding to this minimum  $m_A$ value is the physics target, $(\Delta g_d/g_d^{SM}) ^{target}$, for the measurement of the coupling. We plot $(\Delta g_d/g_d^{SM})^{target}$ as a function of $\tan \beta$ in Fig.~\ref{fig:rdt}. We  also vary $\eta$ because, as already discussed, a non-zero $\eta$ can have a significant effect on the coupling deviations. We find that substantial deviations are possible for low values of $\tan \beta$ because the minimum value of $m_A$, for which only one Higgs boson can be seen with  300 fb$^{-1}$ data, is low in this case (for $\tan \beta=5$ this value is $m_A=200$ GeV).  For $\eta=0$ we get   small deviations for large $\tan \beta$. The $\eta=0$ case is important  because, as previously explained, in the interesting cases with  no mixing and maximal mixing we have $\eta\approx 0$. If  the superpartners are heavy and inaccesible, it would correspond to the no-mixing scenario.


The above was a semi-analytic, semi-model-independent analysis of supersymmetric Higgs coupling deviations under various radiative corrections scenarios. We wish now to investigate the MSSM numerically with a few well-motivated assumptions about the spectrum. 
The sfermion soft breaking diagonal mass parameters have been chosen to be $M_{\text{SUSY}} = 1.2$~TeV for all sfermions except for top squark parameters which will be varied and the smuon mass parameters, which are assumed as $M_{\text{SUSY}}/3$ in order to fulfill the 
constraints for $g-2$~\cite{gminus2} more easily, and even allow favorable contributions at large 
$\tan\beta$ to explain the observed deviation with respect to the SM prediction.  
It should be noted that increasing $M_{\text{SUSY}}$  does not change the results 
significantly.  The trilinear couplings for all the sfermions except for the top squarks are $A_f = 500$~GeV. The gaugino mass parameter  has the value $M_2 = 500$~GeV and $M_1$ is related to $M_2$ via the GUT relation.  The gluino mass is $m_{\tilde{g}} = 1.1$~TeV.  

For effects from the Higgs boson sector, besides the parameters $m_A$ and $\tan \beta$, the most relevant parameters are those entering the top sector, the diagonal soft 
breaking mass parameter $M_{L_{\tilde{Q}_3}}$, $M_{R_{\tilde{t}}}$, the top squark mixing $X_t$, and  additionally $\mu$. These parameters have been scanned: $m_A$ from 200 to 800~GeV, $\tan \beta$ from 2 to 45, $M_{L_{\tilde{Q}_3}} = M_{R_{\tilde{t}}}$ from 100 to 3000~GeV, 
$\mu$ between $\pm 1000$~GeV and $X_t$ between $\pm 150~\text{GeV} \cdot n_{\text{max}}$ 
where $n_{\text{max}}$ is the nearest smaller integer to
$2 M_{L_{\tilde{Q}_3}}/ 150~\text{GeV}$. 

For the scan the program {\tt FeynHiggs2.8.6} \cite{FeynHiggs} has been used for the calculation of the Higgs boson masses, where the lighter CP-even Higgs boson is required to have a mass of $123~\text{GeV} \le  m_h \le 127$~GeV, the Higgs boson couplings and the applied 
constraints~\cite{gminus2, MWandSW, bsgamma}. Only points which fulfill the $W$~boson mass and the electroweak mixing angle 
constraints~\cite{ MWandSW} of precision electroweak analysis have been taken into account in the following.

\begin{figure}[t]
\includegraphics[width=0.9\columnwidth]{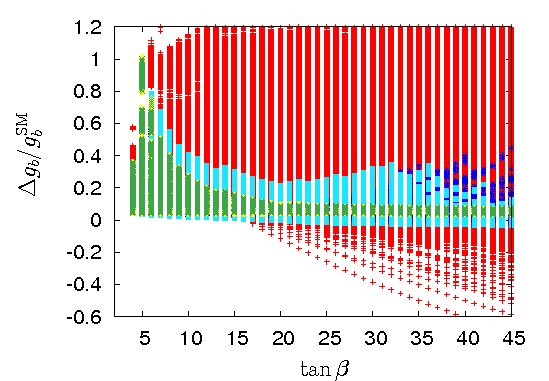}
\caption{
$\Delta g_b/g_b^{\text{SM}}$ as a function of $\tan \beta$. The colour code is the following: Red means several Higgs bosons can be discovered at the LHC - all the other points
 correspond to a single Higgs boson discovery at the LHC. Dark blue points are excluded by 
the $\Gamma(b \rightarrow s \gamma)$ constraint. Light blue, yellow and green correspond to
 at least one third generation~squark has a mass less than 1.0~TeV, all third generation squarks are heavier than 1.0~TeV but at least one top squark is lighter than 1.5~TeV and  both top squarks 
heavier than 1.5~TeV, respectively.}
\label{fig:hbbMSSM}
\end{figure}

\begin{figure}[t]
\includegraphics[width=0.9\columnwidth]{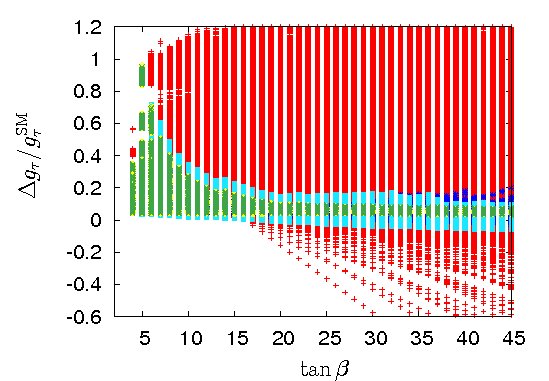}
\caption{
$\Delta g_\tau/g_\tau^{\text{SM}}$ as a function of $\tan \beta$. The colour code is the following: Red means several Higgs bosons can be discovered at the LHC - 
all the other points
 correspond to a single Higgs boson discovery at the LHC. Dark blue points are excluded by 
the $\Gamma(b \rightarrow s \gamma)$ constraint. Light blue, yellow and green correspond to
 at least one third generation~squark has a mass less than 1.0~TeV, all third generation squarks are heavier than 1.0~TeV but at least one top squark is lighter than 1.5~TeV and both top squarks 
heavier than 1.5~TeV, respectively.}
\label{fig:htataMSSM}
\end{figure}

In Fig.~\ref{fig:hbbMSSM} and Fig.~\ref{fig:htataMSSM} the results for $\Delta g_b/g_b^{\text{SM}}$ and $\Delta g_\tau/g_\tau^{\text{SM}}$ are shown, respectively. The red points refer to parameter points where several Higgs bosons will be found according to the corresponding values of $m_A$ and $\tan \beta$ while all the other points represent parameter points where only a single Higgs boson can be discovered.  For the physics target  the latter ones are the ones to focus on. Comparing Fig.~\ref{fig:htataMSSM} with Fig.~\ref{fig:rdt} we find a similar behavior in both plots. Bigger positive deviations can be found for small $\tan \beta$ where there exists the possibility of finding no other Higgs bosons for much lower $m_A$ (see Fig. 1.21 of~\cite{cdr}), which is what allows great volatility in the Higgs couplings when diagonalizing to the mass basis. 
For larger $\tan \beta$ also negative deviation are possible. 

 In Fig.~\ref{fig:hbbMSSM}, for high $\tan \beta$, positive deviations larger than those seen  in the semi-analytical analysis of Fig.~\ref{fig:rdt} or in the $h\tau \tau$ couplings in Fig.~\ref{fig:htataMSSM} are encountered. 
This is due to $\Delta_b$ effects~\cite{Deltab} which are $\tan \beta$ enhanced. $\Delta_b$ effects arise due to  a loop-induced coupling of the Higgs field $H_u$ to the bottom quarks,  $H_u$ being the Higgs  field that couples only to up-type quarks at tree-level. These effects go beyond an effective $\alpha$ approach and are included into the MSSM $hbb$~coupling.  
The effect is small in comparison for the $h \tau \tau$~coupling and is neglected. Requiring that the ratio of  the partial decay branching fraction $B(b \to s\gamma)$~\cite{bsgamma} in the MSSM and the SM lies in the interval of 0.5 and 1.5 leads to the exclusion of part of the parameter points of the single Higgs boson discovery region (dark-blue points), especially also points enhanced by the $\Delta_b$ corrections \footnote{The branching fraction $B(B \rightarrow \mu \mu$) on the other hand, constrains parameter space with small $m_A$ and large $\tan \beta$ which is the parameter region where several Higgs bosons can be found, see eg. Ref.~\cite{Bmumu}.}. This is because the $b\to s\gamma$ amplitude is very similar to the finite $b$-quark mass amplitudes except there is a flavor-changing vertex and a photon attached. Thus, very large $\Delta_b$ is correlated well with an unacceptably large $B(b\to s\gamma)$ deviation.

We now consider the scenario where no superpartners are allowed to be seen at the LHC. Thus, the only phenomenon discovered would be a light Higgs boson.  Among all the superpartners, the third generation squark sector has the most significant impact on Higgs boson coupling deviations (mainly due to 
the large top Yukawa coupling), therefore, we investigate this case by considering various hierarchies of mass limits in this sector. The 
picture will hardly change if the masses of the other superpartners are varied. In Fig.~\ref{fig:hbbMSSM} the light blue is the region where at least one of the third generation squarks is lighter than 1~TeV, yellow is the  region where all the third generation squarks are heavier than 1~TeV but at least one top squark is lighter than 1.5~TeV and the green  region is where  both top squarks are heavier than 1.5~TeV. If one of the light blue parameter points is realized in nature, most likely one of the third generation squarks will be discovered at the LHC. This will be harder for the parameter points corresponding to the yellow region and even more for the green region. The yellow and green regions sit more or less on top of each other. The above criteria for the discovery reach of third generation squarks is coarse, but a more sophisticated criterion, that goes beyond the descriptive needs of our paper, will not make a big qualitative difference.  We find that for large 
$\tan \beta$, the size of the possible deviations, as indicated by the green region, is substantially reduced in case of  heavy top squarks. 
We find typical values of $|\Delta hbb|<5\% - 10\%$ for large $\tan\beta$. As expected, the deviations are much larger for low $\tan\beta\sim 5$ due to the small allowed $m_A$ effect. 

The deviations of the $hVV$ and $htt$ couplings from the SM are as small as expected from the previous discussion. Since $hgg$ and $h\gamma\gamma$ are mostly generated by top quark and $W$ loops, respectively, there is little deviation in these effective couplings also. However, it should be kept in mind that the branching fraction of 
$B(h\to \gamma\gamma)$, or to any Higgs final state for that matter, is affected by the deviation in the $b$-quark Yukawa coupling due to 
$\Gamma(h\to bb)$ comprising a sizable contribution to the denominator of any branching ratio computation.

\textbfx{Conclusions}
We have investigated for a $\sim 125\gev$ SM-like Higgs boson the physics-based targets for its couplings to vector bosons and fermions within three different well-motivated scenarios of physics beyond the SM. The target is based on determining what the possible maximum coupling deviation is  if no other Higgs state is found, nor any non-SM state associated with electroweak symmetry breaking, that would directly reveal that electroweak symmetry breaking is accomplished by something more than a single standard Higgs boson.  The results of the previous discussion are summarized in table~\ref{table:results}. We see that the variation is from less than $1\%$ to over $100\%$ depending on the final state and the new physics scenarios.  The vector boson coupling deviations are expected to be less than $10\%$ in all the cases, and miniscule in the case of supersymmetry. The top quarks coupling is slightly more volatile, and the $b$ quark coupling to the Higgs boson is the most volatile with deviations up to $100\%$ possible within the supersymmetric framework.  The last row in Table~\ref{table:results} reports anticipated $1\sigma$ LHC sensitivities at $14\tev$ with $3\, {\rm ab}^{-1}$ of accumulated luminosity~\cite{Klute:2012pu}.
\begin{table}[t]
\begin{center}
\begin{tabular}{lccc}
\hline\hline 
       & $\Delta hVV$ & $\Delta h\bar tt$ & $\Delta h\bar bb$ \\
\hline
Mixed-in Singlet & 6\% & 6\% & 6\% \\
Composite Higgs & 8\% & tens of \% & tens of \% \\
Minimal Supersymmetry & $<1\%$ & 3\% & $10\%^a$, $100\%^b$ \\ 
LHC $14\tev$, $3\, {\rm ab}^{-1}$ &  $8\%$ & $10\%$ & $15\%$ \\
\hline\hline
\end{tabular}
\caption{\label{table:results} Summary of the physics-based targets for Higgs boson couplings to vector bosons, top quarks, and bottom quarks. The target is based on scenarios where no other exotic electroweak symmetry breaking state (e.g., new Higgs bosons or $\rho$ particle) is found at the LHC except one: the $\sim 125\gev$ SM-like Higgs boson. For the $\Delta h\bar bb$ values of supersymmetry, superscript $a$ refers to the case of high $\tan\beta>20$ { and} no superpartners are found at the LHC, and superscript $b$ refers to all other cases, with the maximum $100\%$ value reached for the special case of $\tan\beta\simeq 5$. The last row reports anticipated $1\sigma$ LHC sensitivities at $14\tev$ with $3\, {\rm ab}^{-1}$ of accumulated luminosity~\cite{Klute:2012pu}.}
\end{center}
\vspace{-0.6cm}
\end{table}

\medskip\noindent
{\it Acknowledgments:} This work is supported in part by the US Department of Energy and by the European Commission under the contract ERC advanced grant 226371 MassTeV. 
We would like to thank S.~Heinemeyer for discussions about {\tt FeynHiggs}  and A.~Thamm for suggestions on the  section on composite Higgs models.



\begin{thebibliography}{99}

\bibitem{ATLAS:2012ae} 
  G.~Aad {\it et al.}  [ATLAS Collaboration],
  Phys.\ Lett.\ B {\bf 710}, 49 (2012)
  [arXiv:1202.1408]; update at ATLAS-CONF-2012-019 (7 March 2012).
  
\bibitem{Chatrchyan:2012tx} 
  S.~Chatrchyan {\it et al.}  [CMS Collaboration],
  arXiv:1202.1488 [hep-ex]; update at CMS-PAS-HIG-12-008 (7 March 2012).

\bibitem{Dittmaier:2011ti}
  S.~Dittmaier {\it et al.}  [LHC Higgs Cross Section Working Group Collaboration],
  arXiv:1101.0593 [hep-ph].

\bibitem{Rauch:2012wa} 
  M.~Rauch,
  arXiv:1203.6826 [hep-ph].
  
\bibitem{Klute:2012pu} 
  M.~Klute, R.~Lafaye, T.~Plehn, M.~Rauch and D.~Zerwas,
  arXiv:1205.2699 [hep-ph].
  
\bibitem{Abe:2010aa} 
  T.~Abe {\it et al.}  [ILD],
  arXiv:1006.3396 [hep-ex].
  
\bibitem{Aihara:2009ad} 
  H.~Aihara {\it et al.} [SiD],
  arXiv:0911.0006 [physics.ins-det].
  
\bibitem{Linssen:2012hp} 
  L.~Linssen {\it et al.} [CLIC],
  arXiv:1202.5940 [physics.ins-det].
  
\bibitem{Martin:1997ns} 
See, e.g., S.~P.~Martin,
  [hep-ph/9709356v6] (6 Sep 2011).
  

\bibitem{Wells}
  R.~Schabinger, J.~D.~Wells,
  Phys.\ Rev.\  {\bf D72}, 093007 (2005)
  [hep-ph/0509209];

\bibitem{Bowen:2007ia}
  M.~Bowen, Y.~Cui, J.~D.~Wells,
  JHEP {\bf 0703}, 036 (2007)
  [hep-ph/0701035].

\bibitem{hep-ph/9504378} 
  M.~Spira, A.~Djouadi, D.~Graudenz and P.~M.~Zerwas,
  Nucl.\ Phys.\ B\ {\bf 453}, 17  (1995)
  [hep-ph/9504378].
\bibitem{mstw} 
  A.~D.~Martin, W.~J.~Stirling, R.~S.~Thorne and G.~Watt,
  Eur.\ Phys.\ J.\ C {\bf 63}, 189 (2009)
  [arXiv:0901.0002 [hep-ph]].
  A.~D.~Martin, W.~J.~Stirling, R.~S.~Thorne and G.~Watt,
  Eur.\ Phys.\ J.\ C {\bf 64}, 653 (2009)
  [arXiv:0905.3531 [hep-ph]].
  \
  A.~D.~Martin, W.~J.~Stirling, R.~S.~Thorne and G.~Watt,
  Eur.\ Phys.\ J.\ C {\bf 70}, 51 (2010)
  [arXiv:1007.2624 [hep-ph]].
  \bibitem{hep-ph/9409380} 
  K.~Hagiwara, S.~Matsumoto, D.~Haidt and C.~S.~Kim,
  Z.\ Phys.\ C\ {\bf 64}, 559  (1994)
  [Erratum-ibid.\ C\ {\bf 68}, 352  (1995)]
  [hep-ph/9409380].

\bibitem{723875} 
  W.~M.~Yao {\it et al.} [Particle Data Group Collaboration],
  J.\ Phys.\ GG\ {\bf 33}, 1  (2006).


 
    \bibitem{ArkaniHamed:2002qy} 
  N.~Arkani-Hamed, A.~G.~Cohen, E.~Katz and A.~E.~Nelson,
  JHEP {\bf 0207}, 034 (2002)
  [hep-ph/0206021].
  \bibitem{Contino:2003ve} 
  R.~Contino, Y.~Nomura and A.~Pomarol,
  Nucl.\ Phys.\ B {\bf 671}, 148 (2003)
  [hep-ph/0306259].
  \bibitem{Agashe:2004rs} 
  K.~Agashe, R.~Contino and A.~Pomarol,
  Nucl.\ Phys.\ B {\bf 719}, 165 (2005)
  [hep-ph/0412089].
  \bibitem{Giudice:2007fh} 
  G.~F.~Giudice, C.~Grojean, A.~Pomarol and R.~Rattazzi,
  JHEP {\bf 0706}, 045 (2007)
  [hep-ph/0703164].
 
  \bibitem{Manohar:1983md} 
  A.~Manohar and H.~Georgi,
  Nucl.\ Phys.\ B {\bf 234}, 189 (1984).
 \bibitem{cdr} 
    CLIC Conceptual Design Report  (2012), see\\http://stanitz.web.cern.ch/stanitz/nightlybuild.html.

\bibitem{Low:2009di} 
  I.~Low, R.~Rattazzi and A.~Vichi,
  JHEP {\bf 1004}, 126 (2010)
  [arXiv:0907.5413 [hep-ph]].
  
  \bibitem{Espinosa:2010vn} 
  J.~R.~Espinosa, C.~Grojean and M.~Muhlleitner,
  JHEP {\bf 1005}, 065 (2010)
  [arXiv:1003.3251 [hep-ph]].
\bibitem{Contino:2010mh} 
  R.~Contino, C.~Grojean, M.~Moretti, F.~Piccinini and R.~Rattazzi,
  JHEP {\bf 1005}, 089 (2010)
  [arXiv:1002.1011 [hep-ph]].
     \bibitem{appear} 
 R.~Contino, C.~Grojean,  D. Pappadopulo, R.~Rattazzi and A. Thamm,
to appear.
\bibitem{Loinaz:1998ph} 
  W.~Loinaz and J.~D.~Wells,
  Phys.\ Lett.\ B {\bf 445}, 178 (1998)
  [hep-ph/9808287].
  
 \bibitem{HaberOthers}
  S.~Heinemeyer, W.~Hollik and G.~Weiglein,
  Eur.\ Phys.\ J.\ C {\bf 16}, 139 (2000)
  [hep-ph/0003022];
  M.~S.~Carena, H.~E.~Haber, H.~E.~Logan and S.~Mrenna,
  Phys.\ Rev.\ D {\bf 65}, 055005 (2002)
  [Erratum-ibid.\ D {\bf 65}, 099902 (2002)]
  [hep-ph/0106116].
  
  
  
  \bibitem{Djouadi:2005gj} 
  A.~Djouadi,
  Phys.\ Rept.\  {\bf 459}, 1 (2008)
  [hep-ph/0503173].
  \bibitem{:1999fr} 
  ATLAS Collaboration (1999)
``ATLAS: Detector and physics performance technical design report. Volume 2,''
  CERN-LHCC-99-15.

\bibitem{gminus2}
  T.~Moroi,
  Phys.\ Rev.\ D {\bf 53} (1996) 6565
   [Erratum-ibid.\ D {\bf 56} (1997) 4424]
  [hep-ph/9512396];
  S.~Heinemeyer, D.~St\"ockinger and G.~Weiglein,
  Nucl.\ Phys.\ B {\bf 690} (2004) 62
  [hep-ph/0312264];
  Nucl.\ Phys.\ B {\bf 699} (2004) 103
  [hep-ph/0405255].

  \bibitem{FeynHiggs} 
S.~Heinemeyer, W.~Hollik and G.~Weiglein,
  Comput.\ Phys.\ Commun.\  {\bf 124} (2000) 76
  [hep-ph/9812320];
  S.~Heinemeyer, W.~Hollik and G.~Weiglein,
  Eur.\ Phys.\ J.\ C {\bf 9} (1999) 343
  [hep-ph/9812472];
  G.~Degrassi, S.~Heinemeyer, W.~Hollik, P.~Slavich and G.~Weiglein,
  Eur.\ Phys.\ J.\ C {\bf 28} (2003) 133
  [hep-ph/0212020];
  M.~Frank, T.~Hahn, S.~Heinemeyer, W.~Hollik, H.~Rzehak and G.~Weiglein,
  JHEP {\bf 0702} (2007) 047
  [hep-ph/0611326];
  K.~E.~Williams, H.~Rzehak and G.~Weiglein,
  Eur.\ Phys.\ J.\ C {\bf 71} (2011) 1669
  [arXiv:1103.1335 [hep-ph]];
see {\tt www.feynhiggs.de} .

\bibitem{MWandSW}
 M.~Awramik, M.~Czakon, A.~Freitas and G.~Weiglein,
  Phys.\ Rev.\ D {\bf 69} (2004) 053006
  [hep-ph/0311148];
  Phys.\ Rev.\ Lett.\  {\bf 93} (2004) 201805
  [hep-ph/0407317];
  A.~Djouadi, P.~Gambino, S.~Heinemeyer, W.~Hollik, C.~J\"unger and G.~Weiglein,
  Phys.\ Rev.\ Lett.\  {\bf 78} (1997) 3626
  [hep-ph/9612363];
  Phys.\ Rev.\ D {\bf 57} (1998) 4179
  [hep-ph/9710438].
\bibitem{Deltab}
 M.~S.~Carena, D.~Garcia, U.~Nierste and C.~E.~M.~Wagner,
  Nucl.\ Phys.\ B {\bf 577} (2000) 88
  [hep-ph/9912516];
 L.~Hofer, U.~Nierste and D.~Scherer,
  JHEP {\bf 0910} (2009) 081
  [arXiv:0907.5408 [hep-ph]].
\bibitem{bsgamma} 
M.~S.~Carena, D.~Garcia, U.~Nierste and C.~E.~M.~Wagner,
  Phys.\ Lett.\ B {\bf 499} (2001) 141
  [hep-ph/0010003];
  T.~Hahn, W.~Hollik, J.~I.~Illana and S.~Penaranda,
  hep-ph/0512315.


  \bibitem{Bmumu} F.~Mahmoudi, S.~Neshatpour and J.~Orloff,
  JHEP {\bf 1208} (2012) 092
  [arXiv:1205.1845 [hep-ph]];
A.~Arbey, M.~Battaglia, A.~Djouadi and F.~Mahmoudi,
  arXiv:1207.1348 [hep-ph].
\end{thebibliography}
\end{document}